\documentclass[aps,prl,preprint]{revtex4}
\usepackage{CJK}

\usepackage{mathtext}
\usepackage{graphicx}

\def\actaa{Acta Astronomica}
\newcommand{\refitem}[5]{\item[]{#1} #2%
\def\REFARG{#3}\ifx\REFARG\TYLDA\else, {\it#3}\fi
\def\REFARG{#4}\ifx\REFARG\TYLDA\else, {\bf#4}\fi
\def\REFARG{#5}\ifx\REFARG\TYLDA\else, {#5}\fi.}

\begin{document}

\begin{CJK*}{GB}{} 
\title{Resident space object detection method based on the connection between Fourier spectrum of the video data difference frame and the linear velocity projection}

\author{V.S.~Baranova}
\author{A.A.~Spiridonov}
\author{V.A.~Cherny}
\author{D.V.~Ushakov}
\author{V.A.~Saetchnikov}
\affiliation{Faculty of RadioPhysics and Computer Technologies, Belarusian State University, Nezavisimosti Avenue 4, Minsk 220030, Belarus, e-mail: vbaranova@bsu.by}

\maketitle
\end{CJK*}

\section{Abstract} 
A method for resident space object (RSO) detection in video stream processing using a set of matched filters has been proposed.  Matched filters are constructed based on the connection between the Fourier spectrum shape of the difference frame and the magnitude of the linear velocity projection onto the observation plane. Experimental data were obtained using the mobile optical surveillance system for low-orbit space objects. The detection problem in testing mode was solved for raw video data with intensity signals from three different satellites: KORONAS-FOTON, CUSAT 2/FALCON 9, GENESIS 1. Difference frames of video data with the AQUA satellite pass to construct matched filters were used. The satellites were automatically detected at points where the difference in the value of their linear velocity projection and the reference satellite was close in value. It has been established that the difference in the inclination angle between the detected satellite intensity signal Fourier image and the reference satellite mask corresponds to the difference in the inclinations of these objects. The proposed method allows not only to detect but also to study the motion parameters of both artificial and natural space objects, such as satellites, debris and asteroids.

Keywords: optical observation, video data, angular measurement, resident space object, Fourier spectrum, difference frame, correlation pattern, linear velocity, orbital determination.

\section{Introduction}
Active space exploration programs entail the overpopulating of the most required orbits by artificial objects of various applications (Blake 2022). According to current data, continuously updated public catalogues include over 50,000 artificial space objects with a size of 10 cm and above, of which 5,677 are operational satellites (Boley \textit{et al.}, Xiangxu Lei \textit{et al.}, Shakun \textit{et al.} 2021, 2018, 2021). Moreover, according to statistical estimates, the total number of functional and non-functional space objects with a diameter of approximately 1 cm in low Earth orbit has already reached one million (Spiridonov \textit{et al.} 2022). Commensurate with the growth in the number of orbital missions, there is a need to scale up existing space surveillance networks, both radio and optical (Pastor \textit{et al.}, Spiridonov \textit{et al.} 2021, 2021). Additional information sources on the space situation using passive optical systems with small aperture telescopes (less than 50 cm) (Danescu \textit{et al.}, Guo, \textit{et al.}, Chun, \textit{et al.} 2022, 2020, 2015) is actively developing and contributes to resolving the problem of congestion in global networks within the scope of regional and national monitoring (Lei, \textit{et al.}, Yanagisawa, \textit{et al.}, Park 2020, 2015, 2018).

Space object detection using optical measurement data is equivalent to the classical problem of detecting a signal in the realization of a stochastic process  against an unknown noise (Stark 1982). The unknown noise is caused by the effects of star twinkling, dynamic light pollution, and atmospheric turbulence instability (Hickson 2018). In general, even the optical signal itself, which represents the reflected light from a space object, is not deterministic. It is necessary to consider that the optical signal from a space object can arise as a random process realization to develop a robust detection method. The uncertainty relates to the intensity modulations due to the heterogeneity of the optical flux reflection  from the space object surface during its orbital rotation. In addition, the uncertainty of the optical signal appears at the moment of the orbit correction, when the space object is moved in some direction relative to an unknown observer. Therefore, it is not enough to analyze individual realizations to determine optimally whether a signal is present in the measured data (Cooke, \textit{et al.}, Oniga, \textit{et al.}, Torteeka, \textit{et al.} 2023, 2018, 2019).

Depending on the optical measurement technology, the space object intensity signal can be presented as an elongated track against a background of stationary stars or an elliptical spot against a background of star tracks(Suthakar, \textit{et al.} 2023). As an evolving concept in existing detection methods that have found practical applications, the convolution of the single-measurement intensity signal or sequence of optical measurements with a morphological filter is used (Chote, \textit{et al.}, Cooke, \textit{et al.}, Sun, \textit{et al.} 2019, 2023, 2015). On the basis of the known waveform, a suitable morphological filter is developed, which detects the maximum correlation when the filter matches the form of the space object intensity (Levesque, Levesque, \textit{et al.} 2009, 2007). Variations in the shape of the morphological filter are limited to synthetic generation of the line, whose length matches the estimated length of the space object track. In (Hickson 2018) the track length assumptions are based on the exposure time and the approximate space object velocity. Another approach to filter development is to generate a line using the coordinates of two points. Point coordinates are calculated using a prediction model based on known space object orbital parameters (Blake, \textit{et al.} 2021). The above approaches use the ideal line of different lengths and orientations as the reference. It is assumed that the space object track in the intensity signal realization will be unambiguously defined. In order for the waveform to be unambiguously defined pre-applied global threshold technology, which removes low-frequency noise and binarizes the image (Wijnen, \textit{et al.}, Diprima, \textit{et al.} 2019, 2018). An evident problem in this case is the uncertainty in the intensity threshold value, which requires an adaptive estimation of the signal-to-noise ratio for each measurement. Convolution with such a morphological filter encounters the problem of ambiguity, where the track length varies depending on the intensity modulation, observation conditions, and the accuracy of the a-priori known orbital parameters. In addition, the full-scale image of the filter is quite a costly computing process.

An alternative to the classical method of optical measurements and the approach to the implementing a matched filter can be the use of video data and features of the difference frame sequence of this data (Baranova, \textit{et al.}, Baranova, \textit{et al.} 2023, 2021). Encoding the optical measurement video stream into a single source file as multiple realizations of a random process allows for accumulating a sufficient amount of information for summation and subtraction operations with minimal noise levels in each individual realization. This representation method depicts a space object as a local area of pixels with varying intensities. The magnitude of displacement within this area depends on the space object velocity, which is accurately determined by its orbital parameters. This implies that there is a functional dependence between the motion dynamics of the object in the frame plane and its orbital parameters. Such functional dependence can be used to extract stable features and build a set of filters in the frequency domain to solve the problem of detecting RSOs using an optical observation system (Spiridonov, \textit{et al.} 2022). 

The paper proposes a method for detecting space objects in video stream processing using a set of matched filters constructed based on the connection between the Fourier spectrum shape of the difference frame and the magnitude of the linear velocity projection onto the observation plane.

\section{Matched-filtering technique for the space objects detection}

The random intensity of the reflected optical flux from the surface space object makes it impossible to establish a-priori values of this parameter as a detection threshold. It follows that the basis for feature extraction should be the texture, or, in other words, the waveform. A single realization of a random process or a single frame is not always sufficient to determine the presence or absence of a signal. Especially when considering a space object as a local pixel region that resembles stars in terms of texture. Therefore, it makes sense to use not just one, but multiple frames of the video stream. The apparent characteristic of a space object in a video sequence is motion. To represent the motion of an object as a distinct texture, it is necessary to use a difference frame (Baranova, \textit{et al.} 2023). The formation of a difference frame is determined by the characteristics of observed space objects: their brightness and velocity. The term RSO  will be used further, which implies the ability to apply the detection model being considered to both artificial and natural space objects, such as satellites, debris and asteroids (Adam{\'o}w, \textit{et al.}, Alton, \textit{et al.}, Wlodarczyk, \textit{et al.} 2022, 2019, 2017).

The Fourier spectrum shape of the difference frame with the intensity signal of the RSO at a given moment in time is closely related to the parameters of its orbital motion. Therefore, given the Fourier spectra of a known RSO with the corresponding angular coordinates and time of observation, it is possible to create a matched filter for detecting another RSO with known or unknown orbital parameters. For this purpose, the dependence of the linear velocity projection onto the frame plane and the declination value within the interval of the pass observation is calculated. Then, reference points with angular coordinates where the Fourier spectrum of the RSO is known on the curve are determined. Based on the known Fourier spectrum, a matched filter is constructed. The matched filter represents a mask with parameters of linear velocity projection and angular coordinate. Sets of matched filters are stored in the local database of the optical observation system as additional data for the catalog of RSO orbital parameters. The flowchart of the detection algorithm is presented in Fig.~1.

\begin{figure}
\centering
\includegraphics[width=.95\textwidth]{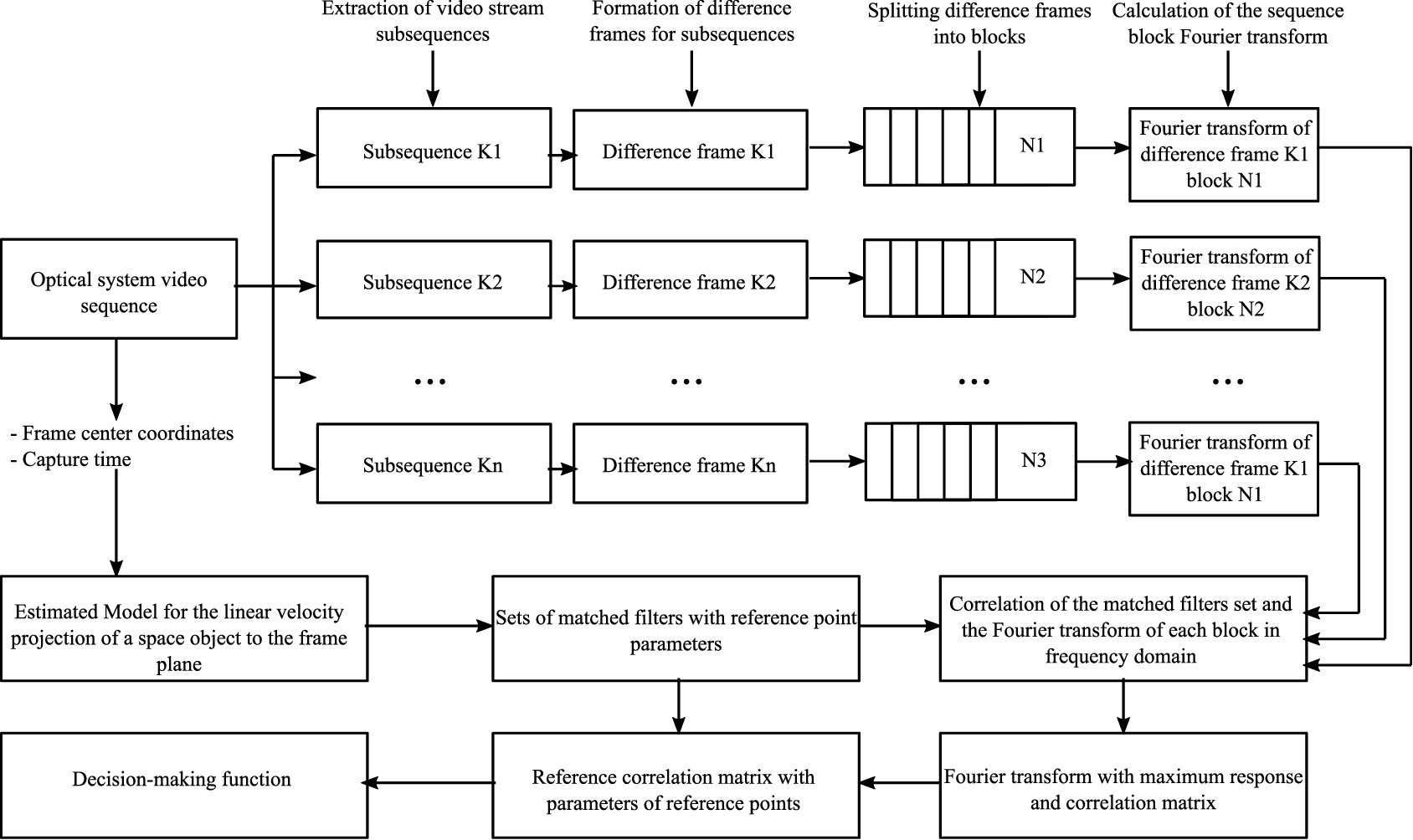}
\caption{Matched-filtering technique for the RSO detection.}
\end{figure}

First, the input video sequence with a total number of frames $N$ is divided into subsequences, where the number of frames in each subsequence is $k$. For each subsequence, a difference frame is generated according to the strategy of subtracting the accumulated sums of n frames. The difference frames of these subsequences are divided into blocks of size ${64}\times{64}$. Then, for each individual block of the difference frame, the Fourier transform is calculated. The obtained Fourier transforms of the input video stream are passed to a part of the processing algorithm, where correlation values between these Fourier transforms and reference masks are calculated in the frequency domain. The correlation of a two-dimensional function $f(x,y)$ of each Fourier transform with a reference mask $h(x,y)$ is determined by the expression:

\begin{equation}
{\mathop{\rm cov}} \left( f\left(x, y\right), h\left(x, y\right) \right) = \sum\limits_{s =  - a}^a {\sum\limits_{t =  - b}^b {h(s,t)} } f(x + s,y + t),
\end{equation}
\noindent
where $x,\, y$ are the spatial variables, ${m}\times{n}$ is the reference mask size, $a=\frac{m-1}{2},\, b=\frac{n-1}{2}$.

According to the two-dimensional convolution theorem (Gonzalez, \textit{et al.} 2008), the inverse discrete Fourier transform of the product of the input signal Fourier transform $F(u,v)$ and the reference mask Fourier transform $H(u,v)$ is equivalent to a two-dimensional convolution $f(x,y)$ and $h(x,y)$ in the spatial domain. The product operation is known to be computationally more efficient than correlation. Therefore, in the detection task, the spectra of the difference frame $F(u,v)$ and reference masks $H(u,v)$ are used in the matched filtering algorithm. For each sequence of difference frames, correlation with a set of matched filters or masks is determined according to the two-dimensional convolution theorem:

\begin{equation}
{\mathop{\rm cov}} (f(x,y),h(x,y)) \Leftrightarrow {F^*}(u,v)H(u,v).
\end{equation}
\noindent
Here, ${F^*}(u,v)$ is complex conjugation of the input signal intensity.

The main equation of matched filtering in the frequency domain, which is used to calculate the correlation pattern $g(x, y)$ in the spatial domain:
\begin{equation}
g(x,y) = {\vartheta ^{ - 1}}[{F^ * }(u,v)H(u,v)],
\end{equation}
\noindent
where $\vartheta ^{ - 1}$ is the inverse Fourier Transform operator.

In a more accurate form, the equation of the correlation pattern in the spatial domain is written taking into account that the Fourier Transform of the difference frame and the reference mask are complex arrays:
\begin{equation}
g(x,y) = {\vartheta ^{ - 1}}[{({F_r}(u,v) + i{F_i}(u,v))^*}({H_r}(u,v) + i{H_i}(u,v)),
\end{equation}
\noindent
where ${F_r}(u,v), \,{F_i}(u,v)$ are real and imaginary parts of the difference frame transform, ${H_r}(u,v), \, {H_i}(u,v)$ are real and imaginary parts of the reference mask transform.

The correlation pattern of difference frame discrete blocks and the reference mask in the detecting RSO algorithm is represented as follows:
\begin{equation}
g(x,y) = abs[{\vartheta ^{ - 1}}[{({F_r}(u,v) + i{F_i}(u,v))^*}({H_r}(u,v) + i{H_i}(u,v))]]{( - 1)^{x + y}}.
\end{equation}

\section{Estimated model for the RSO linear velocity projection to the frame plane}

The estimation model for reference points matched with the known RSO Fourier transform is based on calculating the linear velocity projection onto the frame plane ${V_{plane}}$ as a function of the angular coordinates of declination ${\delta}$. The idea is that for each moment in time where the projections of the RSO linear velocity are the same, the correlation between the spectra of their difference frames will be close to or equal to unity in the case of normalized data. Therefore, the matched filter represents the Fourier spectrum of the known RSO difference frame at a certain point with known values of linear velocity projection and angular coordinates. Such a matched filter will detect the presence of an unknown RSO in the frame plane at points with velocities that are closest to the reference object.

The dependence of the linear velocity projection ${V_{plane}}$ onto the frame plane as a function of declination ${\delta}$ is calculated as follows. The topocentric coordinate system associated with the frame plane is presented in Fig.~2. Based on the known RSO state vectors ${\bf{r}}\left( {X,Y,Z} \right)$ and ${\bf{V}}\left( {{V_X},{V_Y},{V_Z}} \right)$ in the Earth-Centered Inertial (ECI) coordinate system $OXYZ$, for each point of the pass interval at which the measurements were taken, the slant range vector ${{\bf{\rho }}_{{X_1}{Y_1}{Z_1}}}$ and its rate of change $\frac{{d{{\bf{\rho }}_{{X_1}{Y_1}{Z_1}}}}}{{dt}}$ in the topocentric equatorial coordinate system ${{O_1}{X_1}{Y_1}{Z_1}}$ associated with the observation point ${O_1}$ are given by:
\begin{equation}
{{\bf{\rho }}_{{X_1}{Y_1}{Z_1}}} = \left( {\begin{array}{*{20}{c}}
{{\rho _{{X_1}}}}\\
{{\rho _{{Y_1}}}}\\
{{\rho _{{Z_1}}}}
\end{array}} \right) = {\bf{r}} - {{\bf{r}}_{OS}} = \left( {\begin{array}{*{20}{c}}
X\\
Y\\
Z
\end{array}} \right) - \left( {\begin{array}{*{20}{c}}
{{r_\delta }\cos ({\theta _{LST}})}\\
{{r_\delta }\sin ({\theta _{LST}})}\\
{{r_K}}
\end{array}} \right),
\end{equation}

\begin{equation}
\frac{{d{{\bf{\rho }}_{{X_1}{Y_1}{Z_1}}}}}{{dt}} = {\bf{V}} - {{\bf{\omega }}_E} \times {\bf{r}} = \left( {\begin{array}{*{20}{c}}
{\frac{{d{\rho _{{X_1}}}}}{{dt}}}\\
{\frac{{d{\rho _{{Y_1}}}}}{{dt}}}\\
{\frac{{d{\rho _{{Z_1}}}}}{{dt}}}
\end{array}} \right) = \left( {\begin{array}{*{20}{c}}
{{V_X}}\\
{{V_Y}}\\
{{V_Z}}
\end{array}} \right) - \left| {\begin{array}{*{20}{c}}
{\bf{i}}&{\bf{j}}&{\bf{k}}\\
0&0&{{\omega _E}}\\
X&Y&Z
\end{array}} \right|,
\end{equation}
\noindent
where ${{\rho }_{X_1},{\rho }_{Y_1}, {\rho }_{Z_1}}$ are slant range vector ${{\bf{\rho }}_{{X_1}{Y_1}{Z_1}}}$ projections onto the coordinate axes of topocentric equatorial coordinate system (Vallado, \textit{et al.} 2013), ${\theta _{LST}}$ is local sidereal time of the observation point, ${{r_\delta }=({{C_E}+H})\cos ({\phi _{OS}})}$ is radius vector projection of the observation point onto the equatorial plane, $C_E$ is radius of Earth curvature in the meridian, ${\phi _{OS}}$ is the observation point geodetic latitude, ${{r_K }=({{S_E}+H})\sin ({\phi _{OS}})}$is radius vector projection of the observation point onto a plane perpendicular to the equatorial plane, ${S_E}$ is Earth curvature parameter, ${H}$ is the observation point height,  ${\bf{r}}$ is RSO position radius vector in the ECI coordinate system, ${{\bf{r}}_{OS}}$ is position radius vector in the observation point relative to the Earth center, ${\bf{V}}$ is RSO velocity vector in the ECI coordinate system, ${{\bf{\omega }}_E}$ is Earth rotation angular velocity.

Topocentric declination ${\delta}$ and right ascension ${\alpha}$, which characterize the RSO position at moment $t$ relative to the observation point, are calculated as follows:
\begin{equation}
{\delta} = \arcsin \left( {\frac{{{\rho _{{Z_1}}}}}{\rho }} \right),\quad
{\alpha} = \arctan\left( {\frac{{{\rho _{{Y_1}}}}}{{{\rho _{{X_1}}}}}} \right).
\end{equation}

Then, the coordinates and projections of the RSO velocity are determined in a topocentric coordinate system ${{O_1}{X_2}{Y_2}{Z_2}}$ associated with the frame plane - the ${{O_1}{X_2}}$ axis is directed perpendicular to the frame plane, and the ${{O_1}{Y_2}}$ and ${{O_1}{Z_2}}$ axes lie in the frame plane (Fig. 2).

\begin{figure}[htb]
\centering
\includegraphics[width=.75\textwidth]{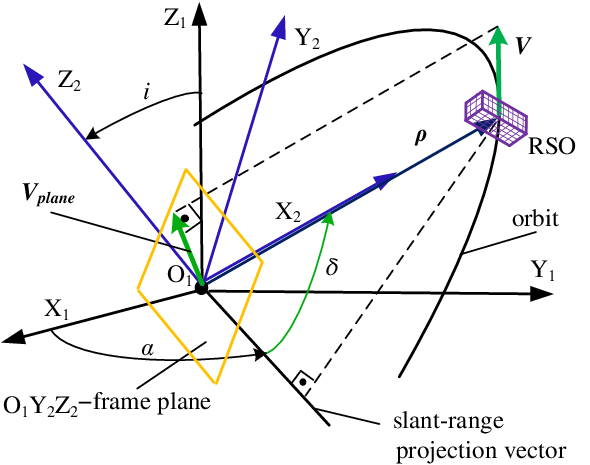}
\caption{Topocentric coordinate system ${{O_1}{X_2}{Y_2}{Z_2}}$ associated with the frame plane.}
\end{figure}

The RSO slant range vector ${{\bf{\rho }}_{{X_2}{Y_2}{Z_2}}}$ and its rate of change $\frac{{d{{\bf{\rho }}_{{X_2}{Y_2}{Z_2}}}}}{{dt}}$ in the topocentric coordinate system ${{O_1}{X_2}{Y_2}{Z_2}}$ associated with the frame plane could be written as:
\begin{equation}
{{\bf{\rho }}_{{X_2}{Y_2}{Z_2}}} = \left( {\begin{array}{*{20}{c}}
{{\rho _{{X_2}}}}\\
{{\rho _{{Y_2}}}}\\
{{\rho _{{Z_2}}}}
\end{array}} \right) = {R_3}\left( \delta  \right) \cdot {R_1}\left( i \right) \cdot {R_3}\left( \alpha  \right) \cdot \left( {\begin{array}{*{20}{c}}
{{\rho _{{X_1}}}}\\
{{\rho _{{Y_1}}}}\\
{{\rho _{{Z_1}}}}
\end{array}} \right),
\end{equation}

\begin{equation}
\frac{{d{{\bf{\rho }}_{{X_2}{Y_2}{Z_2}}}}}{{dt}} = \left( {\begin{array}{*{20}{c}}
{\frac{{d{\rho _{{X_2}}}}}{{dt}}}\\
{\frac{{d{\rho _{{Y_2}}}}}{{dt}}}\\
{\frac{{d{\rho _{{Z_2}}}}}{{dt}}}
\end{array}} \right) = {R_3}\left( \delta  \right) \cdot {R_1}\left( i \right) \cdot {R_3}\left( \alpha  \right) \cdot \left( {\begin{array}{*{20}{c}}
{\frac{{d{\rho _{{X_1}}}}}{{dt}}}\\
{\frac{{d{\rho _{{Y_1}}}}}{{dt}}}\\
{\frac{{d{\rho _{{Z_1}}}}}{{dt}}}
\end{array}} \right).
\end{equation}

Rotation matrices that rotate a vector around the  ${{O_1}{X_1}}$ and ${{O_1}{Z_1}}$ axes are given by:
\begin{equation}
{R_1}\left( i \right) = \left( {\begin{array}{*{20}{c}}
1&0&0\\
0&{\cos \left( i \right)}&{\sin \left( i \right)}\\
0&{ - \sin \left( i \right)}&{\cos \left( i \right)}
\end{array}} \right),\quad
{R_3}\left( \alpha  \right) = \left( {\begin{array}{*{20}{c}}
{\cos \left( \alpha  \right)}&{\sin \left( \alpha  \right)}&0\\
{ - \sin \left( \alpha  \right)}&{\cos \left( \alpha  \right)}&0\\
0&0&1
\end{array}} \right).
\end{equation}

Then the projection of the RSO linear velocity ${V_{plane}}$ onto the frame plane at time $t$ is determined by the following formula:
\begin{equation}
{V_{plane}} = \sqrt {{{\left( {\frac{{d{\rho _{{Y_2}}}}}{{dt}}} \right)}^2} + {{\left( {\frac{{d{\rho _{{Z_2}}}}}{{dt}}} \right)}^2}}.
\end{equation}

\section{Experimental details}
The possibility of RSO detection in video data using the matched filtering method was investigated. Experimental data were obtained by the mobile optical surveillance system for low-orbit space objects at the Belarusian State University (Spiridonov, \textit{et al.}, Baranova, \textit{et al.} 2022, 2023 ). The optical system with an effective field of view $8^{\circ}{14^{''}}$ based on the wide-angle apochromatic lens RedCat 51 APO with a focal length of 250 mm and the f-number 4.9 is mounted on a computerized rotator with the maximum slew rate of $4^{\circ}/{s}$. The sensor is a full-format Canon EOS R camera with a video data transfer rate via the HDMI port of up to 480 Mbps. The frame size is in pixels ${1920}\times{1080}$ with an.avi encoding format. The optical surveillance system for low-orbit space objects provides orbit angular measurements of operational satellites and space debris objects up to 9th magnitude. The mobile optical system is shown in Fig.~3~(left). An example of a difference frame of a 10 s video sequence with the AQUA satellite intensity signal is shown in Fig.~3~(right).
\begin{figure}[htb]
\centering
\includegraphics[width=.32\textwidth]{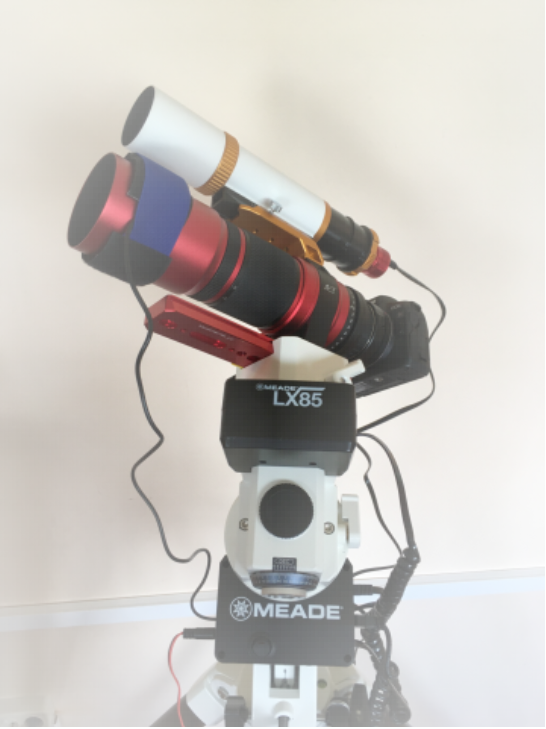}
\includegraphics[width=.57\textwidth]{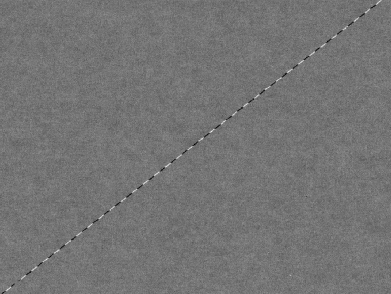}
\caption{Left: mobile optical surveillance system. Right: difference frame of the AQUA satellite intensity signal.}
\end{figure}

To construct a matched filter, difference frames of video data with the AQUA (NORAD:27424) satellite pass were used at three capture intervals in a region with known angular coordinates. Preliminary results from processing experimental videos with intensity signals from various satellites have shown that the Fourier spectra of the differential frames obtained by subtracting the accumulated sums of every three frames exhibit the highest uniformity and stable signal-to-noise ratio. Therefore, in most experiments, a difference frame calculated according to the following formula was used as the basis for the intensity signal characterizing the dynamics of the RSO motion:

\begin{equation}
\Delta f(x,y,{t_k}) = \left| {\Delta f(x,y,{t_{k - 1}}) - \sum\limits_{i = 0}^n {f(x,y,{t_{i + 3}})} } \right|,
\end{equation}

\begin{equation}
\Delta f(x,y,{t_{k-1}}) =  {\left|{\sum\limits_{i = 0}^n {f(x,y,{t_{i}})} - \sum\limits_{i = 0}^n {f(x,y,{t_{i - 3}})} }\right| } ,
\end{equation}
\noindent
where $\Delta f(x,y,{t_k})$ is $k$-th difference frame of the video sequence, $\Delta f(x,y,{t_{k-1}})$ is previous difference frame of the video sequence, ${f(x,y,{t_{i}})}$ is $i$-th frame of the video sequence, ${n=2}$ is number of frames for accumulating the sum.

\section{Results and discussions}

\subsection{Detection of satellite intensity signal by matched-filtering technique}
The detection problem was solved for raw video data with intensity signals of three satellites taken from a series of observations for 05.08.2022-06.08.2022: KORONAS-FOTON(NORAD:33504), CUSAT 2/FALCON 9(NORAD:39271), GENESIS 1(NORAD:29252). The orbital parameters of these satellites and the reference satellite AQUA, are presented in Table 1. Here, $r_p$ and $r_a$ are perigee and apogee radius, $i$ is orbit inclination and $T$ indicates the orbit period.

\begin{table}[!t]
\caption{Orbital parameters of the satellites.}
\begin{tabular}{*{7}c}
\hline
No&Sattelite&NORAD ID&	$r_p$, km&	$r_a$, km&	$i$, deg&	$T$, min\\ \hline
0&AQUA&	27424	&705.1&	706.3&	98.3&	98.7 \\ \hline
1&KORONAS-FOTON&	33504&	495.6&	520.8	&82.4&	94.6\\ \hline
2&GENESIS 1&	29252&	477.9&	495.1	&64.5&	94.2\\ \hline
3&CUSAT 2/FALCON 9&	39271&	308.5	&881.5&	80.9&	96.4\\ \hline
\end{tabular}
\end{table}

The functional dependence of the RSO linear velocity projection ${V_{plane}}$ onto the frame plane from the declination ${\delta}$ was calculated for each satellite under consideration in relation to the dependence of the reference satellite AQUA ${\delta_{r}}$.

As a result of the matched filtering algorithm, the satellites KORONAS-FOTON, CUSAT 2/FALCON 9, GENESIS 1 were automatically detected in a sequence of difference frame blocks in the region with the following angular coordinates: KORONAS-FOTON- ${\alpha _{1}}=18h14m17s$, ${\delta_{1}=31^{\circ}{01^{'}}{52^{''}}}$, CUSAT 2/FALCON 9- ${\alpha _{3}}=16h24m51s$, ${\delta_{3}=69^{\circ}{45^{'}}{36^{''}}}$, GENESIS- ${\alpha _{2}}=05h18m56s$, ${\delta_{2}=71^{\circ}{40^{'}}{42^{''}}}$. 

According to the functional dependence obtained from the linear velocity projection onto the frame plane from the declination, space objects were detected at the corresponding points using masks that contained the intensity signal of the AQUA satellite with linear velocity projection values close to the detected objects. The calculated curves of the linear velocity projection onto the frame plane from the declination with marked points at which objects were detected are shown in Fig.4. Additionally, points of the curve for the reference object are marked, where the generated masks identified a correlation pattern with the maximum coefficient values. In the Fig.4 reference and detected objects curve points marked with identical markers: AQUA and KORONAS-FOTON points are red and black circles, for AQUA and GENESIS points marked by black and green squares, AQUA and CUSAT 2/FALCON 9 points are black and blue triangles. The resulting correlation patterns for each pair of points of the reference mask and the detected satellite, in particular, AQUA and KORONAS-FOTON~(Correlation pattern 1), AQUA and GENESIS~(Correlation pattern 2), AQUA and CUSAT 2/FALCON 9~(Correlation pattern 3) are presented in Fig.~5, respectively. The reconstructed areas of difference frames with the intensity signal of the detected satellites, masks of the AQUA satellite at reference points and their Fourier images are shown in Fig.~6.

\begin{figure}
\centering
\includegraphics[width=.85\textwidth]{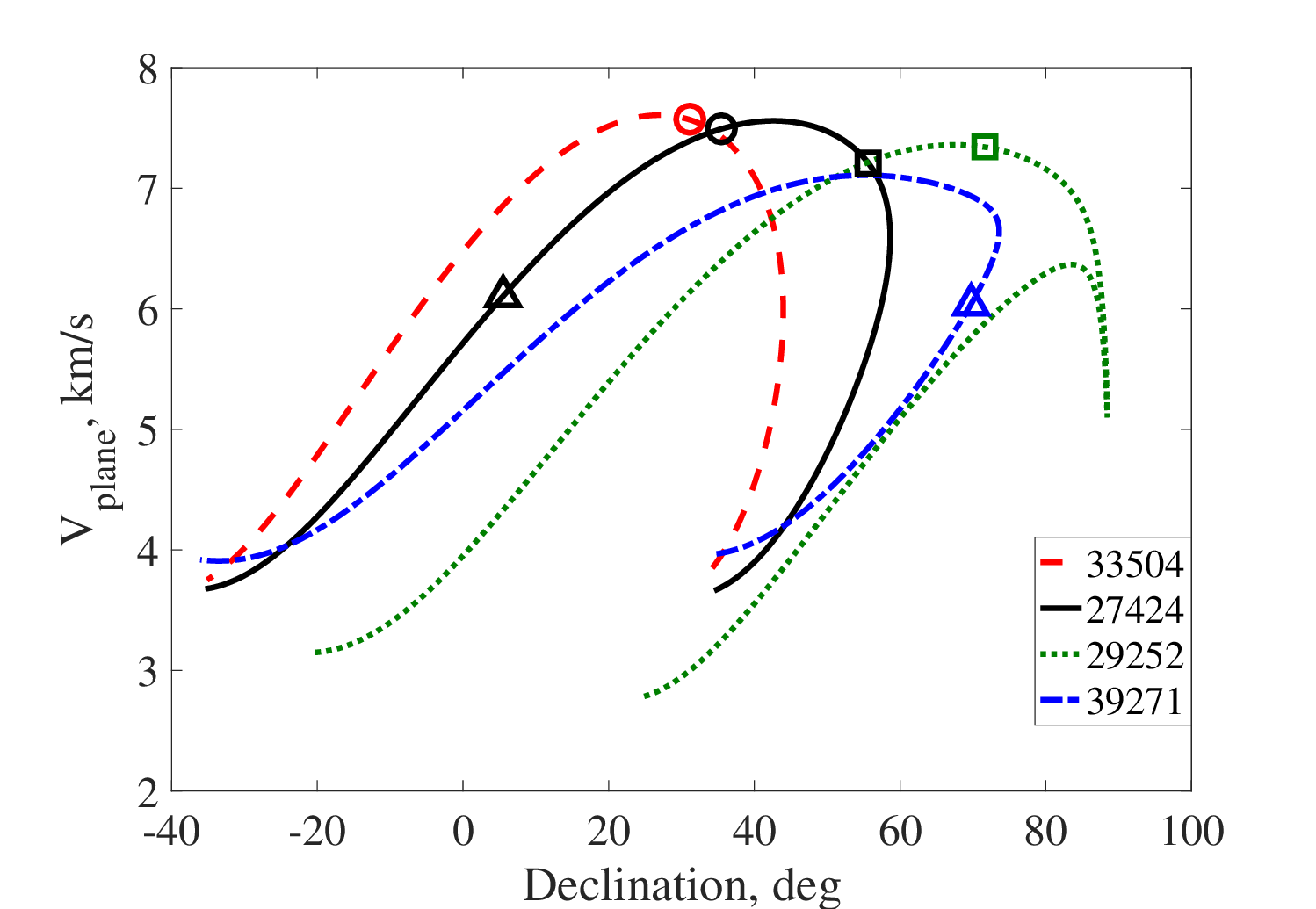}
\caption{Curves of the linear velocity projection onto the frame plane from the declination of space objects AQUA (27424), KORONAS-FOTON (33504), GENESIS (29252), CUSAT 2/FALCON 9 (39271).
\vspace{0.2cm}}
\centering
\includegraphics[width=.31\textwidth]{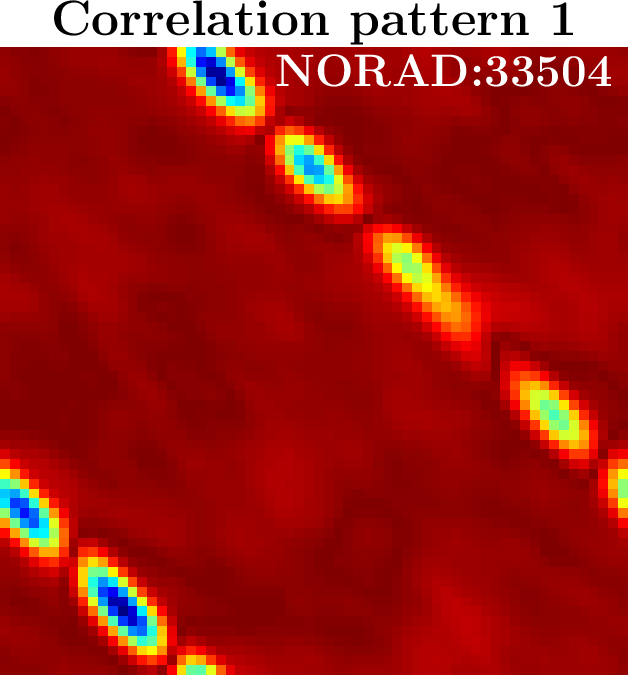}
\includegraphics[width=.31\textwidth]{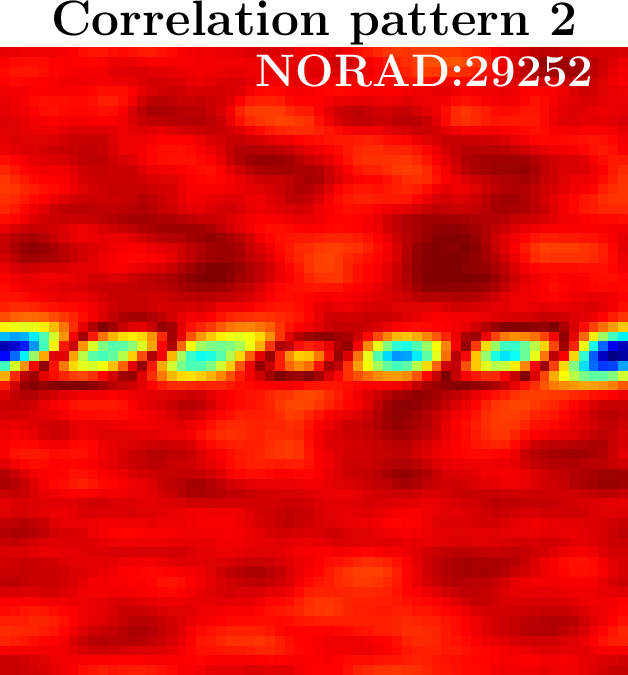}
\includegraphics[width=.31\textwidth]{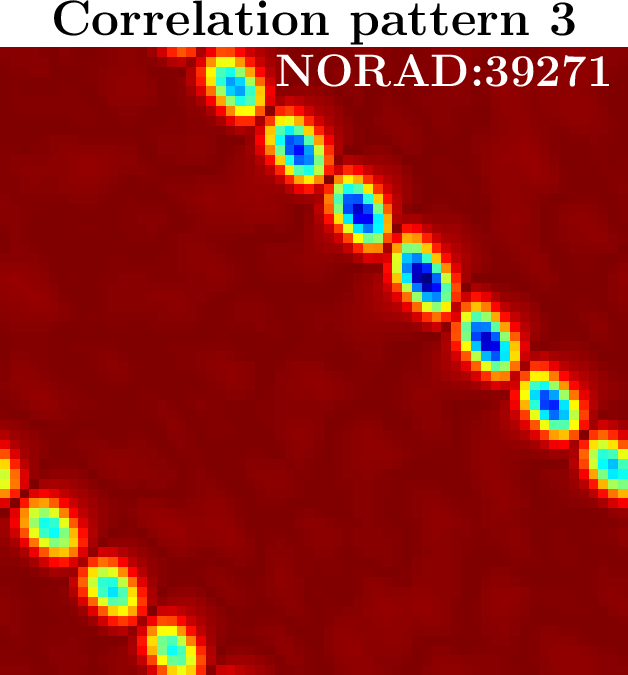}
\caption{Correlation patterns for points pairs of the reference mask and the detected satellite.}
\end{figure}

Analyzing the obtained results, it can be concluded that the closer the values of the linear velocity projection onto the frame plane of the reference mask and the detected satellite, the higher the maximum correlation value and the symmetry of the correlation patterns. For the CUSAT 2/FALCON 9 satellite, the curve of the linear velocity projection ${V_{plane}}$ from the declination ${\delta}$ (Fig. 4, blue line with triangle marker) reflects the minimum difference between the value ${V_{plane}}$ at declination $\delta_{3}=69^{\circ}{45^{'}}{36^{''}}$ and the value ${V_{plane}}$ of the generated mask with AQUA satellite (Fig. 4, black line with triangle marker) at declination $\delta_{r_{3}}=05^{\circ}{27^{'}}{46^{''}}$ is equal to $\Delta V_{plane_{3}}=0.062$~km/s. The correlation image of the reference satellite AQUA and the detected satellite CUSAT 2/FALCON 9 is presented in~Fig.~5~(right).

Notably, the correlation pattern for this satellite contains an additional component that corresponds to the frequency component of the Fourier image of the differential frame. This is caused by the inhomogeneity of the intensity signal of both the satellite itself and the reference object mask. This effect is observed in the Fourier images of difference frames in Fig.~6 for the CUSAT 2/FALCON 9 satellite and the reference mask with the AQUA satellite. As is the case for the detected satellite KORONAS-FOTON.

\begin{figure}
\centering
\includegraphics[width=.23\textwidth]{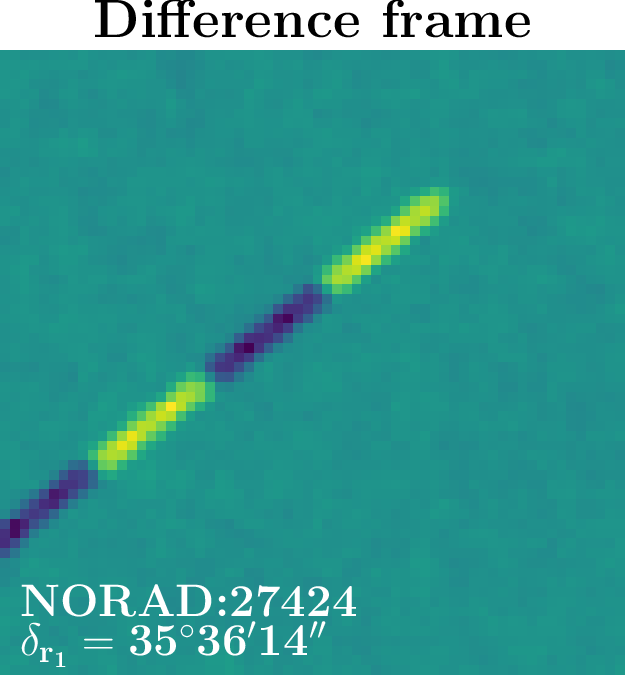}
\includegraphics[width=.23\textwidth]{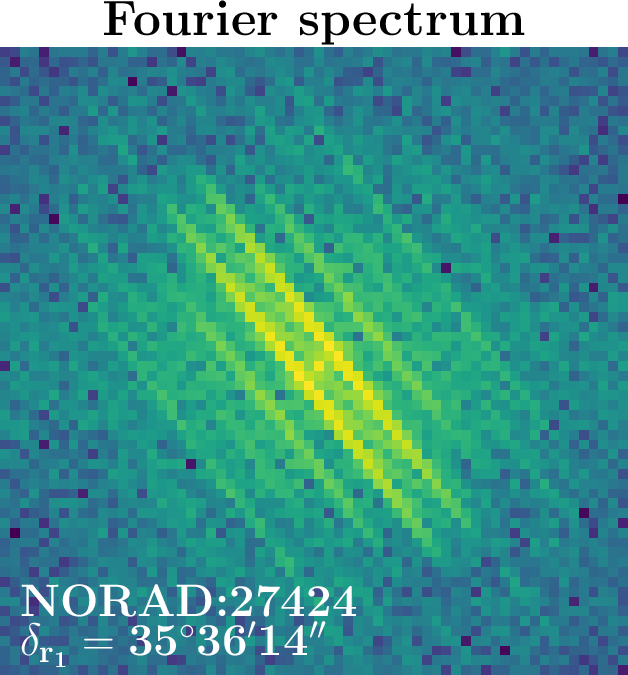}
\includegraphics[width=.23\textwidth]{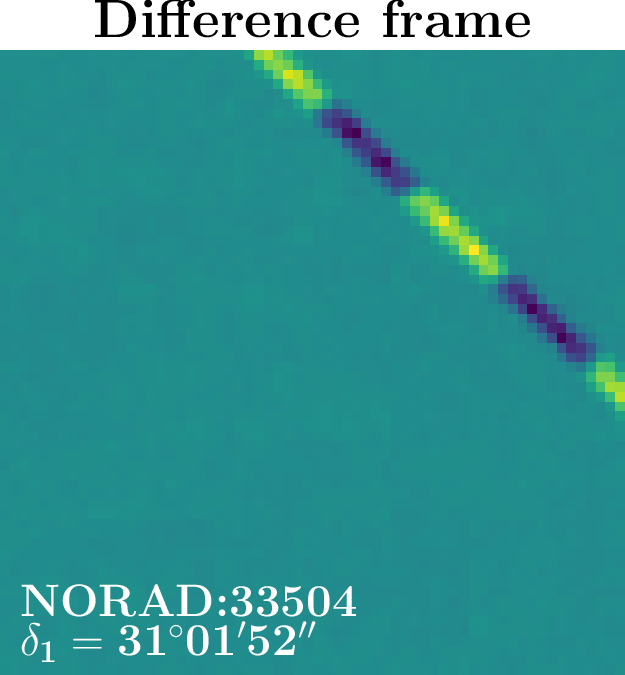}
\includegraphics[width=.23\textwidth]{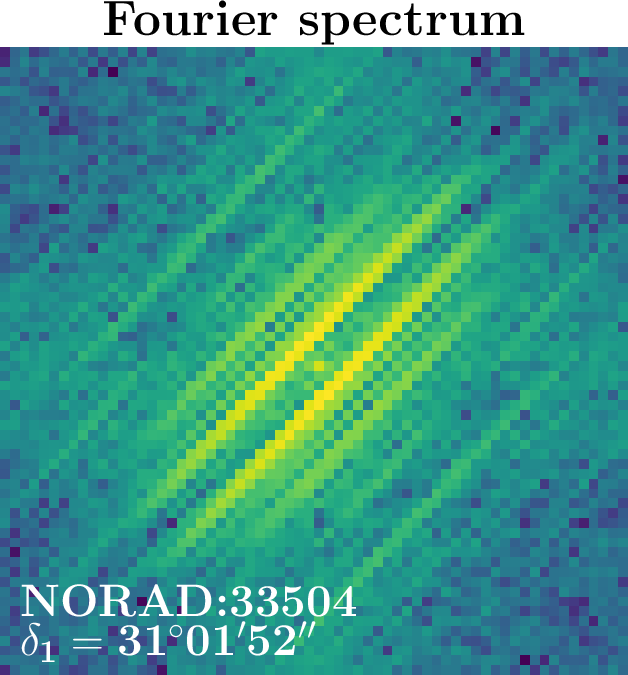}
\includegraphics[width=.23\textwidth]{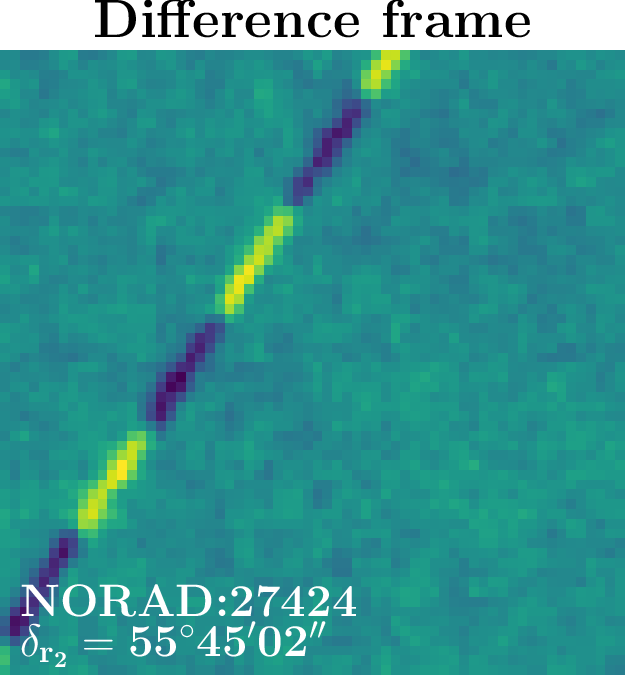}
\includegraphics[width=.23\textwidth]{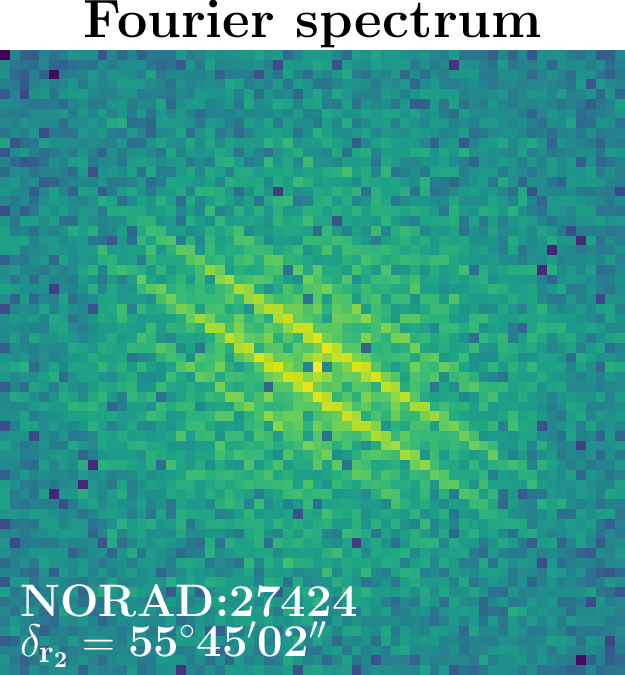}
\includegraphics[width=.23\textwidth]{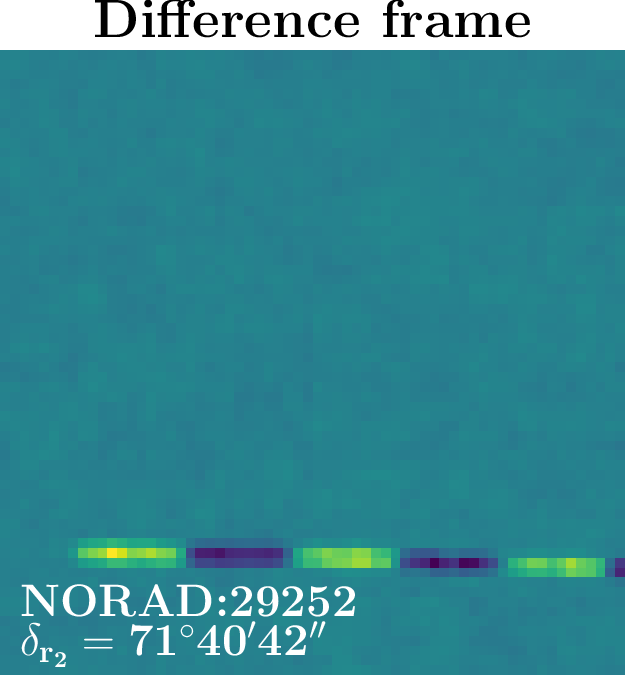}
\includegraphics[width=.23\textwidth]{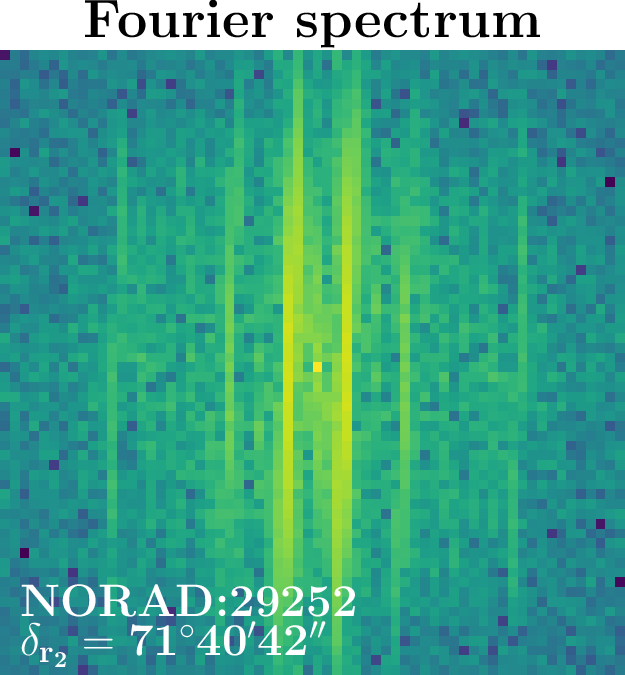}
\includegraphics[width=.23\textwidth]{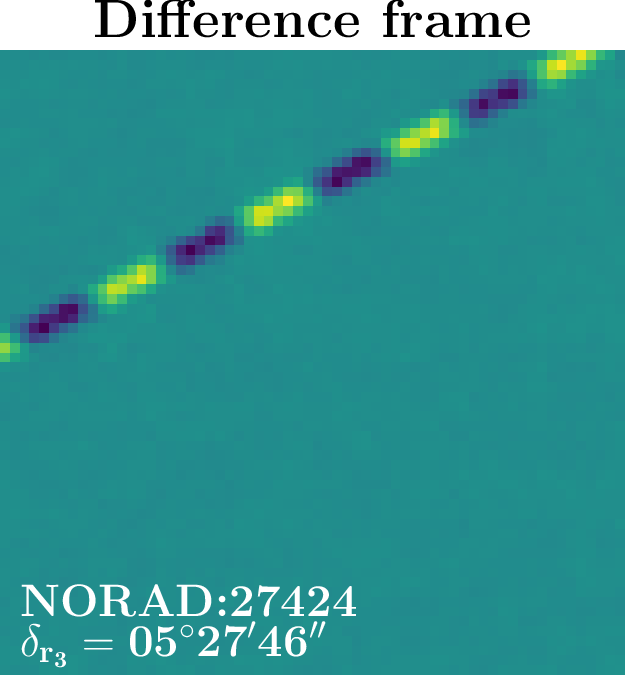}
\includegraphics[width=.23\textwidth]{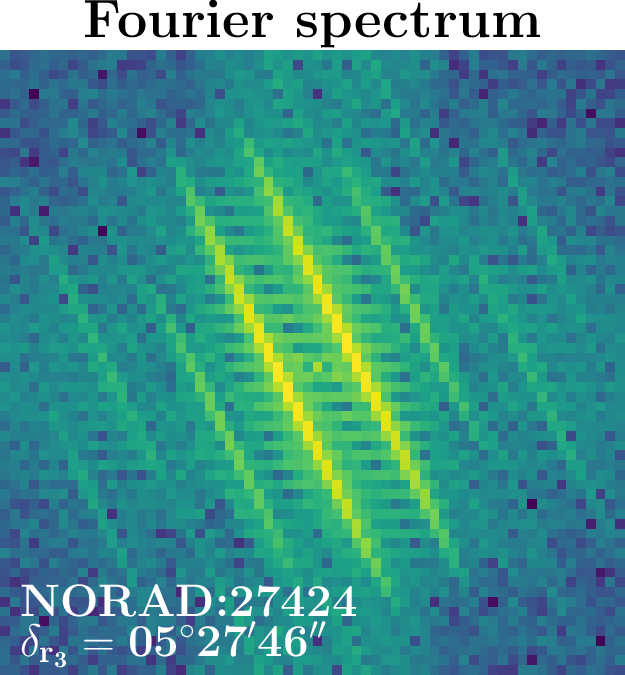}
\includegraphics[width=.23\textwidth]{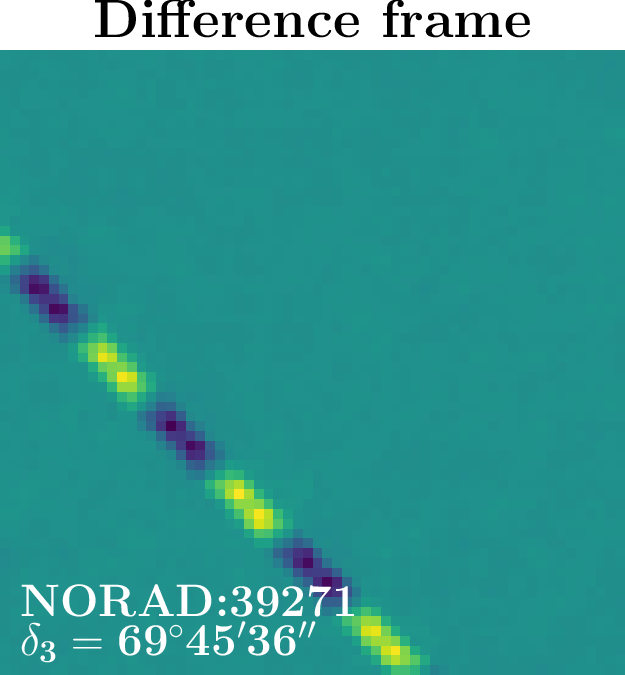}
\includegraphics[width=.23\textwidth]{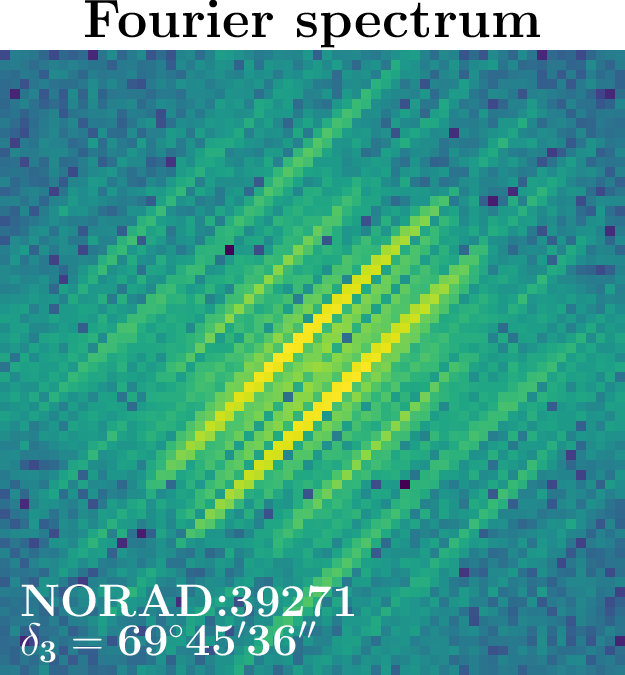}
\caption{Reconstructed areas of difference frames with the detected satellites intensity signal and the AQUA satellite at reference points (circles, triangles and squares in accordance with Fig.~4) and their Fourier images.}
\end{figure}

For the KORONAS-FOTON satellite, the correlation pattern of which is shown in Fig.~5~(left), the curve of the linear velocity projection ${V_{plane}}$ from the declination ${\delta_{t}}$~(Fig.~4, red line with circle marker) shows the difference of ${V_{plane}}$ in the point region ${\delta_{1}=31^{\circ}{01^{'}}{52^{''}}}$ and the value of ${V_{plane}}$ in the declination region ${\delta_{r_{1}}=35^{\circ}{36^{'}}{14^{''}}}$ of the generated AQUA satellite mask~(Fig.~4, black line with circle marker) is equal to $\Delta V_{plane_{1}}=0.077$~km/s. Accordingly, for the detected GENESIS satellite, the difference between the projection of linear velocity ${V_{plane}}$ at declination ${\delta_{2}=71^{\circ}{40^{'}}{42^{''}}}$(Fig. 4, green line with square marker) and the value ${V_{plane}}$ of the generated mask with AQUA satellite~(Fig.~4, black line with square marker) at declination ${\delta_{r_{2}}=55^{\circ}{45^{'}}{02^{''}}}$ is $\Delta V_{plane_{2}}=0.137$~km/s. The correlation pattern of AQUA and GENESIS is presented in Fig.~5~(middle).

It is worth noting that the difference in the inclination angle of the Fourier image of the reconstructed intensity signal of the detected satellites and the reference satellite mask with preliminary mirror operation corresponds to the difference in the orbit inclinations of these objects (Fig.~6). This is confirmed by the obtained mask rotation angle values for the maximum correlation value with detected objects: ${\Delta{\phi}=16^{\circ}}$ for AQUA and KORONAS-FOTON, ${\Delta{\phi}=34^{\circ}}$ for AQUA and GENESIS, as well as ${\Delta{\phi}=17^{\circ}}$ for AQUA and CUSAT 2/FALCON 9. The established dependence will be investigated in further work.

\section{Conclusions}
A method for detecting RSOs in video data using a set of matched filters was proposed. Matched filters are constructed based on the connection between the Fourier spectrum shape of the difference frame and the magnitude of the linear velocity projection onto the observation plane. The research was carried out as part of the stable detection system development for mobile optical surveillance station of low-orbit space objects at the Belarusian State University.

To construct a matched filter, difference frames of video data with the AQUA satellite pass were used at three capture intervals in a region with known angular coordinates. The detection problem was solved for raw video data with intensity signals from three satellites: KORONAS-FOTON, CUSAT 2/FALCON 9, GENESIS 1. As a result of the matched filtering algorithm, the satellites KORONAS-FOTON, CUSAT 2/FALCON 9, GENESIS 1 were automatically detected in a sequence of difference frame blocks in the region with the following angular coordinates: KORONAS-FOTON- ${\alpha _{1}}=18h14m17s$, ${\delta_{1}=31^{\circ}{01^{'}}{52^{''}}}$, CUSAT 2/FALCON 9- ${\alpha _{3}}=16h24m51s$, ${\delta_{3}=69^{\circ}{45^{'}}{36^{''}}}$, GENESIS- ${\alpha _{2}}=05h18m56s$, $\delta_{2}=71^{\circ}{40^{'}}{42^{''}}$.

The functional dependence of the RSO linear velocity projection ${V_{plane}}$ onto the frame plane from the declination ${\delta_{t}}$ was calculated for each satellite under consideration in relation to the dependence of the reference satellite AQUA. Correlation patterns were obtained for each point pair of the reference mask and the detected satellite, specifically, AQUA and KORONAS-FOTON, AQUA and GENESIS, as well as AQUA and CUSAT 2/FALCON 9. The satellites were detected at points where the difference in the value of their linear velocity projection and the reference satellite was: AQUA and KORONAS-FOTON $\Delta V_{plane_{1}}=0.077$~km/s, AQUA and GENESIS $\Delta V_{plane_{2}}=0.137$~km/s, AQUA and CUSAT 2/FALCON 9 $\Delta V_{plane_{3}}=0.062$~km/s.

It has been established that the difference in the inclination angle of the Fourier image of the reconstructed intensity signal of the detected satellites and the reference satellite mask with preliminary mirror operation corresponds to the difference in the orbit inclinations of these objects. This is confirmed by the obtained mask rotation angle values for the maximum correlation value with detected objects: ${\Delta{\phi_{1}}=16^{\circ}}$ for AQUA and KORONAS-FOTON, ${\Delta{\phi_{2}}=34^{\circ}}$ for AQUA and GENESIS, as well as ${\Delta{\phi_{3}}=17^{\circ}}$ for AQUA and CUSAT 2/FALCON 9.

The proposed method allows not only for detection but also for studying the motion parameters of RSOs. In particular, the Fourier image of the difference frame provides an approximation of the velocity, the angular coordinate of declination, and the satellite orbit inclination relative to the reference space object. These parameters serve as the basis for constructing the matched filter.

\section*{Acknowledgements}
This work was supported by the Republic of Belarus scientific research State programs "High-tech technologies and equipment" and "Digital and space technologies, human, society and state security".


\begin{references}
\refitem{Adam{\'o}w, M., Niedzielski, A.~T., Wolszczan, A., Maciejewski, G.}{\actaa}{2022}{72}{257-266}
\refitem{Alton, K.~B., St{{e}}pie{\'n}, K.}{2019}{\actaa}{69}{283-304}
\refitem{Baranova, V., Spiridonov, A.,  Liashkevich, S., Saetchnikov, V.}{2023}{Proc. IEEE 10th Int. Workshop on Metrology for AeroSpace}{~}{551}
\refitem{Baranova, V.S., Saetchnikov, V.A.,  Spiridonov, A.A.}{2021}{Devices and Methods of Measurements}{12}{272}
\refitem{Blake, J.}{2022}{Astronomy and Geophysics}{63}{2.14}
\refitem{Blake, J.A., Chote, P.,  Pollacco, D.,  Feline, W.,  Privett, G.,  Ash, A., Eves, S.  et al.}{2021}{Adv. Space Res.}{67}{360}
\refitem{Boley, A.C., Byers, M.}{2021}{Sci. Rep.}{11}{10642}
\refitem{Chote, P.,  Blake, J.A., Pollacco, D.}{2019}{Adv. Maui Optical and Space Surveillance Technologies Conf.}{~}{52} 
\refitem{Chun, F., Tippets, R., Della-Rose, D. J., Polsgrove, D., Gresham, K., and Barnaby, D. A.}{2015}{American Astronomical Society Meeting Abstracts}{225}{424}
\refitem{Cooke, B.F.,  Chote, P.,   Pollacco, D., West, R., Blake, J.A.,  McCormac, J., Airey, R., Shrive, B.}{2023}{Adv. Space Res.}{72}{907}
\refitem{Danescu, R.G., Itu, R., Muresan, M.P., Rednic, A., Turcu, V.}{2022}{Remote Sensing}{14}{1905}
\refitem{Diprima, F., Santoni, F.,  Piergentili, F.,  Fortunato, V., Abbattista, C.,  Amoruso L.}{2018}{Acta Astronautica}{145}{332}
\refitem{Gonzalez, R.C., Woods, R.E.}{2008} {Digital image processing}{~}{~}
\refitem{Guo, X., Gao, P., Shen, M., Yang, D., Yu, H., Liu, T., Li, J., Zhao, Y.}{2020}{Adv. Space Res.}{65}{1990}
\refitem{Hickson, P.}{2018}{Adv. Space Res.}{62}{3078}
\refitem{Lei, X.; Li, Z.; Du, J.; Chen, J.; Sang, J.; Liu, C.}{2020}{Adv. Space Res.}{67}{350}
\refitem{Levesque, M.}{2009}{Proc. of the Adv. Maui Opt. and Space Surveillance Technologies Conf.}{~}{E81}
\refitem{Levesque, M.P.,  Sylvie, B.}{2007}{Technical rept}{~}{~} 
\refitem{Oniga, G.,  Miron, M.,  Danescu, R. and  Nedevschi, S.}{2011}{IEEE 7th Int. Conf. on Intelligent Computer Communication and Processing}{~}{335}
\refitem{Park, J.-H.}{2018}{Adv. Space Res.}{62}{152} 
\refitem{Pastor, A., Sanjurjo-Rivo, M., and Escobar, D.}{2021}{Adv. Space Res.}{68}{2677}
\refitem{Shakun, L., Koshkin, N., Korobeynikova, E., Kozhukhov, D., Kozhukhov, O., and Strakhova, S.}{2021}{Adv. Space Res.}{67}{1743}
\refitem{Spiridonov A.A., Baranova V.S., Saetchnikov V.A., Ushakov D.V.}{2022}{J. of the Belarusian State University. Physics.}{2}{50}
\refitem{Spiridonov, A., Baranova, V.,  Ushakov, D., Saetchnikov, V.,  Kenko, Z.,  Zasmuzhats, D., and  Mechinsky, V.}{2022}{Proc. IEEE 9th Int. Workshop on Metrology for AeroSpace}{~}{566}
\refitem{Spiridonov, A., Saetchnikov, V.,  Ushakov, D.,  Cherny, V.,  Kesik, A.}{2021}{Proc. IEEE 8th Int. Workshop on Metrology for AeroSpace}{~}{132}
\refitem{Spiridonov, A.A.,  Kesik, A.G., Saetchnikov, V.A.,   Cherny, V.E., Ushakov, D.V.}{2021}{IEEE J. on Miniaturization for Air and Space Systems}{2}{59}
\refitem{Stark, H.}{1982}{Application of Optical Fourier Transforms}{~}{~}
\refitem{Sun, R.-y.,  Zhan, J.-w., Zhao, C.-y.,  Zhang, X.-x.}{2015}{Acta Astronautica}{110}{9}
\refitem{Suthakar, V., Sanvido, A.A., Qashoa, R., Lee, R.S.K. }{2023}{Sensors}{23}{9668}
\refitem{Torteeka, P., Gao, P.-q.,  Shen, M.,  Guo, X.-z., Yang, D.-t.,  Yu, H.-h., Zhou, W.-p.,  Tong, L., and  Zhao, Y.}{2019}{Adv. Space Res.}{63}{461} 
\refitem{Vallado, D.A., McClain, W.D.}{2013} {Fundamentals of Astrodynamics and Applications}{~}{~}

\refitem{Wijnen, T.,  Stuik, R., Rodenhuis, M.,  Langbroek, M.,  Wijnja, P.}{2019}{Proc. 1st NEO and Debris Detection Conf.}{~}{437}
\refitem{Wlodarczyk, I., {\v{C}}ernis, K., Boyle, R.~P.}{2017}{\actaa}{67}{81}
\refitem{Xiangxu Lei, Kunpeng Wang, Pin Zhang, Teng Pan, Huaifeng Li, Jizhang Sang, Donglei He}{2018}{Adv. Space Res.}{62}{542}
\refitem{Yanagisawa, T., Kurosaki, H., Oda, H., and Tagawa, M.}{2015}{Adv. Space Res.}{56}{414}

\end{references}
\end{document}